\documentclass[12pt]{article}

\usepackage[top=2.72cm, bottom=2.08cm, left=2.61cm, right=2.61cm]{geometry}  
\usepackage{mathptmx}      
\usepackage[T1]{fontenc}   
\usepackage[utf8]{inputenc}
\usepackage{microtype}     
\usepackage[labelfont=bf]{caption} 

\usepackage{graphicx}      
\usepackage{booktabs}      
\usepackage{float}         
\usepackage{setspace}      
\usepackage{rotating}      

\usepackage[none]{hyphenat}
\usepackage[english]{babel}  

\usepackage{amsfonts}      
\usepackage{nicefrac}      

\usepackage{natbib}        
\usepackage{doi}           
\usepackage{url}           
\usepackage{hyperref}      

\usepackage{authblk}       

\usepackage[hang, flushmargin]{footmisc}
\newcommand{\keywords}[1]{%
  \vspace{0.5em}
  \noindent\textbf{Keywords: }#1
  \vspace{1em}
}

\title{In the Network of the Conclave: Social Connections and the Making of a Pope}

\author[a,b]{Giuseppe Soda}
\author[a,b]{Alessandro Iorio}
\author[a,b,c]{Leonardo Rizzo}

\affil[a]{Bocconi University, Department of Management of Technology, Via Roentgen 1, 20136, Milan, Italy}
\affil[b]{SDA Bocconi, Network Innovation Lab, Via Sarfatti 10, 20136, Milan, Italy}
\affil[c]{Central European University, Department of Network and Data Science, Quellenstraße 51, 1100, Wien, Austria}

\date{}

\begin{document}
\maketitle

\begin{abstract}
\noindent \normalsize
This study brings a network perspective to papal elections by mapping the relational architecture of the College of Cardinals. Using publicly available data sources, such as official Vatican directories and episcopal consecration records, we assemble a multiplex network that captures cardinals’ co-membership in various collegial bodies of the Vatican and their consecration ties. We then calculate structural metrics to capture three key mechanisms that we suggest have a crucial role in the dynamics of the conclave: status, mediation power, and coalition building. Our descriptive study—publicly released prior to the May 8, 2025 election of Pope Leo XIV—shows that Cardinal Robert F. Prevost, largely ignored by pundits, bookmakers, and AI models, held a uniquely advantageous position in the Vatican network, by virtue of being central in multiple respects. Thus, although being considered an underdog by many, the network perspective suggests that Cardinal Prevost was de facto one of the strongest “papabile.”
\end{abstract}

\pagebreak

\section{Introduction}
\begin{quote}
\singlespacing
\textit{Q: \textquotedblleft Is the Holy Spirit responsible for the
election of a pope?\textquotedblright }\\
\textit{A: \textquotedblleft I would not say so, in the sense that
the Holy Spirit picks out the Pope\ldots{} I would say that the Spirit
does not exactly take control of the affair, but rather, like a good
educator, as it were, leaves us much space, much freedom, without
entirely abandoning us. Thus the Spirit\textquoteright s role should
be understood in a much more elastic sense---not that he dictates
the candidate for whom one must vote.}\textit{\textquotedblright{}}

(Interview with Cardinal Ratzinger, later Pope Benedictus XVI)
\end{quote}
Few decision-making processes in the modern world remain as cloaked in rituals, secrecy, and symbolism as the papal conclave \citep{allen2002conclave}. When the College of Cardinals gathers in the Sistine Chapel to elect a new pope, the world watches for the white smoke that signals the outcome of a process often described as guided by divine inspiration. Yet, as the quote above suggests, the process is not so much steered by divine forces as it is shaped by human judgment and negotiation. The fact that the Holy Spirit does not indicate which candidate one should vote for—as Cardinal Ratzinger stated—reveals that even in theological debate, it is recognized that the conclave’s decision-making process, like in many other human organizations, is not immune to mechanisms such as mediation power, coalition building, and status dynamics. Although the cardinals in the conclave are secluded from the rest of the world, they are not isolated from the force field of social influences among them built through interpersonal connections \citep{padgett1993robust,krackhardt1990assessing}. Indeed, we argue that the conclave is a complex, highly relational decision-making process, where influence flows through network ties that form the social scaffolding through which consensus can be built \citep{urena2019social,stevenson2000agency}. These ties create a complex network that connects the cardinals, allowing them to get to know each other, work together, exchange ideas, and transfer information \citep{granovetter1973strength,reagans2003network}. We ask, therefore: What role does the network—and the structural position each cardinal occupies within it—play in shaping the likelihood that someone can emerge as frontrunner candidate during the conclave?

The question is as consequential as it is difficult to answer. Obtaining reliable data to understand why some cardinals emerge as key figures and how the college converges toward a candidate—and, more generally, what happens during the conclave—is, in fact, challenging, if not impossible. Those who participate in the conclave are bound to absolute silence, even years after the fact, under penalty of excommunication.\footnote{The Apostolic Constitution \textit{Universi Dominici Gregis} (John Paul II, 22 February 1996), which sets forth the norms for the conduct of the conclave, establishes the penalty of \textit{latae sententiae} excommunication: “for those involved in the conclave who are bound to secrecy, in the event of violating that secrecy (n. 58)”.}

We apply social network theories and methods to enhance the understanding of the papal conclave by examining how interpersonal connections among cardinals may shape decision-making \citep{borgatti2009network}. This relational perspective allows us to move beyond anecdotal interpretations, offering a systematic lens for studying how patterns of connections may influence voting behavior. While certain ties—such as trust or support—may emerge or evolve during the conclave, the backbone of the network structure, built through long-standing institutional, personal, and professional relationships, originates prior to it. Analyzing the preexisting networks, therefore, provides a vantage point for anticipating the dynamics of consensus formation within the conclave. Specifically, while acknowledging that formal ecclesiastical hierarchies define roles and responsibilities, we suggest that influence and consensus dynamics within the conclave are embedded in relations, which can be made visible and analytically tractable through a network perspective. In doing so, we can model the conclave as a multiplex social system, and thus illuminate key network mechanisms underlying papal succession.

Drawing on publicly available data, such as official Vatican documents, books, and journalistic reports, we systematically reconstruct the network leading up to the 2025 papal conclave to identify which individuals occupy central, bridging, or otherwise structurally advantageous positions within the College of Cardinals. In other words, we provide a systematic, network-based method for identifying potential \textit{papabili}.\footnote{\textit{Papabile} (pope-able), plural \textit{papabili}, is an unofficial Italian term coined by Vaticanologists and used internationally in many languages to describe a Catholic man—in practice, always a cardinal—who is thought of as a likely or possible candidate to be elected pope by the College of Cardinals. This term has become so pervasive that is commonly used in the society to indicate all possible candidates to any type of succession.}  Using standard tools from social network analysis—including centrality measures and community detection algorithms—we evaluate each cardinal’s structural position within the ecclesiastical network \citep{freeman1977set,podolny2005status}.

Importantly, this was not an ex-post exercise. Our model and its results were finalized and shared prior to the white smoke that signaled the election of Pope Leo XIV on May 8, 2025. On May 7, 2025, we completed our analyses and shared the results on Corriere della Sera and Bocconi University’s website. That same evening, Corriere della Sera prepared a full article based on our findings, which went to print overnight. By the morning of May 8\textsuperscript{th}, hours before the outcome of the conclave, hard copies of the newspaper featuring our network-based ranking of \textit{papabili} were available at newsstands across Italy (see Figure A1 in the Appendix for the printed version of the article). In this sense, our model did not just interpret events after the fact. Rather, it provided a structural analysis that entered the public domain before the final outcome was released.

Our approach is rooted entirely in network theory and mathematics, taking advantage of a decades of research on network status, information control and mediation power, and coalition building \citep{podolny2001networks,burt1992structural,mizruchi1992structure}. We did not leverage natural language processing, deep learning, or other artificial intelligence tools. Rather, this is a case of applying a classic social network analysis approach to a novel setting—leveraging co-membership ties and apostolic lineages—to shed light on one of the most secretive decision-making processes in the world. Our findings suggest that understanding who is papabile is not only a matter of ideology, but also of relational embeddedness, that is, being known, trusted, and connected in the right ways to the right others \citep{useem1986inner,heinz1993hollow,krackhardt1993informal}.

\section{The Networks Behind the Pyramid}
\label{sec:pyramid}
\subsection{Collegial Structures}
By virtue of its millennial resilience, unique among human organizations, the Roman Catholic Church is often cited as a paradigmatic example of a top-down organizational hierarchy \citep{coopman2000power}. In fact, its internal order is defined by a clear chain of authority \citep{noll2022turning}. Canon law and ecclesiastical tradition assign specific offices—starting from the pope and Roman Curia down through cardinals, bishops, priests, and deacons—each with formally defined duties, so that tasks and responsibilities are divided, coordinated, and supervised through an official hierarchy. However, on much more careful and rigorous analysis, this appears to be a simplistic, if not stereotyped, view of how such a complex organization actually operates. In several key passages of \textit{The Code of Canon Law} it is stated the “collegial” spirit underlying key decision-making processes within the Church.\footnote{A major reorganization initiated by Pope Pius X was incorporated into the 1917 Code of Canon Law, marking a significant step in formalizing the structure of the Roman Curia. Further reforms were undertaken by Pope Paul VI in the 1960s, aimed at modernizing procedures and increasing the international representation within the Curia. These efforts were reflected in the 1983 revision of the Code of Canon Law. In 1988, Pope John Paul II introduced another wave of structural and functional reforms through the apostolic constitution Pastor bonus (“The Good Shepherd”), which served as the guiding document for curial operations for over three decades. This was eventually superseded in 2022 by a new apostolic constitution, Praedicate evangelium (“Preach the Gospel”), issued by Pope Francis, which further redefined the Curia’s mission and organization.}  For instance Can. 353 §1 clarifies that “Cardinals assist the Supreme Pastor of the Church in collegial fashion particularly in Consistories...” and Can. 349 established that “[...] The cardinals assist the Roman Pontiff either collegially when they are convoked to deal with questions of major importance, or individually when they help the Roman Pontiff through the various offices they perform, especially in the daily care of the universal Church.” 

Thus, while the hierarchical structure of the Catholic Church reflects a vertical chain of command, with the pope at the top, such a depiction risks obscuring the complex, consultative, and collaboration-based processes that characterize its governance and organization. Central to this collaborative approach is the role of various collegial bodies, such as dicasteries, congregations, commissions, and councils, which support and advise the pope in exercising his office. These include institutions like the Dicastery for the Doctrine of the Faith, the Council for the Economy, and the Synod of Bishops, all of which function with a spirit of shared responsibility and collective discernment.\footnote{Since the Second Vatican Council (1962–1965), there has been a significant theological and organizational emphasis on episcopal collegiality—the principle that, while the Pope holds primacy, bishops collectively share responsibility for governing the universal Church.}  Thus, while final authority rests with the pope, the actual process involves substantial interaction, negotiation, and consensus-building among several actors. Like in many large and complex organizations, also in the Catholic Church the extensive use of group-based decision-making and coordination mechanisms creates an articulated web of interactions among individuals by virtue of shared presence within these groups.

Similar to directors sitting in corporate boards \citep{mizruchi1996interlocks}, or actors collaborating in movies \citep{watts1998collective}, the shared participation of cardinals in the formal collegial bodies of the Roman Curia constitute ties in the co-membership networks \citep{breiger1974duality}. Thus, to capture and analyze how structural influence unfolds among cardinals, we used a well-established technique from social network analysis by projecting a bipartite network—linking cardinals to collegial organizational bodies—into a cardinal-by-cardinal co-membership network \citep{pfeffer2007everything}. While our context differs, we draw methodological inspiration from \citet{porter2005network}, who applied a similar approach to study co-membership patterns among legislators in the U.S. House of Representatives. In our case, the two-mode network connects cardinals to the collegial bodies to which they are assigned. Through collaboration within these bodies, cardinals cultivate mutual familiarity, exchange ideas, information, expertise, and provide reciprocal support. To summarize, the network-based lens highlights that decision-making in the Catholic Church—especially at the Vatican level—is shaped through complex webs of interlocking memberships and collaborative structures. This network originating from formal assignments constitutes a fundamental component of the salient relational infrastructure relevant to the conclave.

To achieve a more fine-grained understanding of the network that connects those entrusted with electing the pope, formal ties should be complemented by more informal connections—such as friendship, advice, or trust \citep{krackhardt1992strength}. Although our study could not reliably capture informal relationships of this kind for all cardinals, we reconstructed another type of tie that enriches the co-membership structure and helps illuminate the multiple social spaces that characterize the conclave’s decision-making process.\footnote{For a small subset of cardinals, we were able to supplement our data with journalistic reports suggesting the presence of informal ties (e.g., personal friendships or long-standing collaborations). We ran our analyses both with and without these additional ties and found no meaningful differences in the results. However, because these data are not systematically available across the full sample, we do not include them in the reported network.}  Specifically, we focus on lines of episcopal consecration, which we describe in detail in the following section.

\subsection{Consecration Lines}
Episcopal consecration inscribes each bishop into an unbroken apostolic lineage, sacramentally linking him to his principal consecrator and, ultimately, to the Apostles. This occurs through a solemn liturgical rite in which an already-consecrated bishop ordains a new bishop by laying hands on him, invoking the Holy Spirit, and reciting the consecratory prayer. This lineage is not merely symbolic—it establishes tangible institutional, theological, and social bonds. Consecration not only confers sacramental authority, but also forges a personal relationship between the consecrator and the ordained—one that can be understood as a strong tie in social network terms \citep{granovetter1973strength}. Theologically, the principal consecrator is considered a “spiritual father,” a role that carries expectations of pastoral care, guidance, and trust \citep{o2000trent}.

Importantly, the rite of consecration typically involves not only the principal consecrator, but also one or two co-consecrators, reinforcing the collegial and communal character of the episcopate. When, for instance, a cardinal consecrates a bishop who is later elevated to the College of Cardinals, a tie is established that extends beyond liturgical function and becomes part of the lived social structure of the Church’s leadership. These consecration ties, meticulously recorded in ecclesiastical annals, reflect the Catholic principle that the Church’s mission is transmitted person-to-person, generation-to-generation, through the sacrament of orders.

In the context of the conclave, these ties take on a concrete political and relational significance. A cardinal who has consecrated another member of the college—whether as principal or co-consecrator—has played a formative role in that person’s ecclesiastical trajectory.  This relationship may imply a sense of loyalty, spiritual indebtedness, or alignment of vision \citep{yuengert2001bishops}. As such, consecration lines create a relational infrastructure of influence and obligation within the electoral body, shaping how support, trust, and alignment might unfold during the papal election.

\section{Network Mechanisms and the Conclave}
\label{sec:mechanisms}
The papal conclave is the formal process through which the Roman Catholic Church elects its new pope. It is one of the oldest and most secretive electoral institutions still in use today, governed by centuries of tradition, detailed canonical legislation, and intricate ritual. Participation in the conclave is restricted to cardinal electors, that is, members of the College of cardinals who are under the age of 80 at the time the papacy becomes vacant. These electors gather in the Sistine Chapel, locked away \textit{cum clave} ("with a key"), to deliberate and vote in complete seclusion from the outside world. 

Despite its highly codified procedures, the conclave is far from a static or predictable event. One key reason is that the actual voting preferences and dynamics remain entirely confidential, even after the end of the conclave.\footnote{Allegedly, the only conclave on which there is information on the voting patterns over different rounds is the 2013 conclave that ended with the election of Cardinal Bergoglio as Pope Francis. This information comes from a book wrote by \citet{o2019election}.}  Ballots are burned, and no official record of individual votes is ever released. This secrecy makes each conclave a unique and analytically challenging decision-making episode. It is reasonable to expect that electors may enter the conclave with a short list of preferred candidates—often shaped by theological affinity, regional interests, curial loyalties, pastoral vision or other personal qualities—but once voting begins, the electoral landscape can shift rapidly. The process is iterative: ballots are cast twice each morning and twice each afternoon, with the requirement of a two-thirds majority for a valid election. Early rounds may serve as signaling devices, allowing electors to test the waters, coalesce around emerging names, or block others \citep{grandori2013handbook}. As such, researchers lack the empirical data needed to retrospectively analyze voting behavior or train predictive models based on past outcomes.

To interpret the dynamics of the conclave that lead to the election of a pope, popular wisdom has produced a well-known saying: “He who enters the conclave as pope leaves as a cardinal.” This adage encapsulates the historical inaccuracy of predictions, which are typically based on the individual attributes of cardinals. While such characteristics may serve as cues for voters \citep{berger1972status}, we theorize that the process by which cardinals become credible frontrunners is fundamentally shaped by relational dynamics. The conclave exemplifies a closed decision-making network, where each participant is both an actor with individual preferences and a node embedded in overlapping layers of influence—co-membership in dicasteries, shared consecration ties, doctrinal alignment, and regional blocs. This social structure extends beyond the cardinals admitted to the conclave itself reaching a broader set of key actors, who are part of the general congregations, consisting in preparatory meetings held in the days before the actual conclave. These meetings offer all cardinals—not only the electors—an opportunity to deliberate and prepare before voting. Unlike the conclave, which is highly codified in both procedures and rituals, the general congregations are less formalized, creating a space where existing social structures shape interaction patterns, mediation efforts, channels of influence, and the formation of preliminary agreements. As such, the boundaries of the network of relationships among cardinals that we investigate goes beyond the cardinal electors.

Understanding the conclave through the lens of social networks helps illuminate not only the institutional structure of the Church, but also the relational mechanics that drive one of the most consequential leadership selections in the modern world. The social network structure of the College of Cardinals provides the scaffolding for the emergence of papal frontrunners. 

Building on this idea, we theorize that the contours of formal organizational boundaries, which in our context include relationships for coordination and resource exchange within collegial bodies, circumscribe the opportunity set for interactions to occur and develop. This concept aligns with a fundamental intuition in social network research regarding the importance of “opportunities” for social interaction, as popularized by \citeauthor{blau2018crosscutting} (1997: 29): “...rates of social association depend on opportunities for social contact”. Within the context of organizations, as Brass and colleagues suggested (\citeyear{brass2004taking}), people’s and groups’ formal positions constrain “...their opportunity to interact with some others and [facilitate] interaction with still others” (\citeyear{brass2004taking}: 796). For instance, employees are more likely to interact if they share an organizational focus such as the same business unit, job function, or workspace \citep{kleinbaum2013discretion}. Thus, the design of organizational structures and processes creates the social infrastructure through which people interact. Beyond the concrete opportunities for interactions offered by the formally designed patterns, there’s also a cognitive effect generated when actors coordinate among each other within the same organizational unit. Specifically, research showed that organizational units identify discursive entities or “zones of meaning that are linguistically circumscribed” \citeauthor{tasselli2020bridging} (2020: 1294). Therefore, membership in groups within an organization, entails both adopting and influencing the shared interpretations that individuals employ to comprehend the organization's social landscape. Thus, formal organizational structures influence organizational actors in developing shared cognitive schemas—particularly among those who belong to the same units—that pertain to various factors: the social structure of the organization, the subgroups within it, the status hierarchy that defines relationships among those subgroups, and the roles individuals play within them \citep{balogun2005intended,podolny2001networks}.

In this context, we suggest that at least three network mechanisms help explain how one candidate may progressively accumulate support and reach the two-thirds majority required for election. The first mechanism is status-based influence, which builds on the principle that connections to highly central individuals confer both visibility and endorsement potential \citep{podolny2005status}. These ties generate advocacy effects: prominent cardinals can signal viability, lend symbolic weight, and help consolidate early votes behind a peer who appears not only competent and trustworthy \citep{soda2024prismatic}. Ties to high-status actors do not merely boost a cardinal’s prestige through association, they actively enhance perceptions of his competence and reliability, thereby increasing his perceived trustworthiness and solidifying the candidacy. Additionally, since the relational infrastructure underpinning the network of cardinals is the outcome of specific organizational choices, prominence within this network also serves as an institutional signal, a normative cue that may further shape voting behavior. Cardinals are aware of each other’s roles, affiliations, and histories and the network in which they are embedded formed by co-membership ties in the collegial bodies and the consecration lines is largely observable from within. In highly symbolic settings such as the conclave, decisions are shaped by normative roles and institutional validation. Under this logic, the candidates’ credibility hinges on whether their affiliations resonate with the institutional logic, that is, the implicit repertoire of what is considered “appropriate” within the conclave’s social and historical context. Thus, what matters is not merely competence in isolation, but whether a candidate’s identity and bearing are perceived as fitting the conclave’s normative fabric—an identity that adheres to the collective understanding of what a pope should embody.

Second, individuals who stand between others in their connections \citep{freeman1977set} may facilitate or hinder the communication and information exchange by occupying positions that allow them to control others’ access to information and resources  \citep{freeman1991centrality}. By virtue of these positions, cardinals can more effectively navigate the social divides without being fully captured by any one group \citep{gargiulo2000trapped}, allowing them to influence perceptions and manage tensions \citep{burt1992structural}. Cardinals occupying these positions can bridge ideological, geographical, or institutional divides without appearing fully aligned with any single faction. This positional flexibility enables such candidates to be perceived as neutral integrators, particularly appealing in politically fragmented or polarized settings \citep{gould1989structures}. Unlike contexts in which mediators and brokers derive value from recombining diverse information to foster innovation \citep{burt2004structural}, the conclave may reward these positions since they help in connecting boundaries by maintaining cohesion across camps. Their ability to interpret, reframe, or transmit messages across possible divides makes them instrumental in fostering consensus. This bridging capacity enhances their symbolic legitimacy as unifying figures, positioning them as "compromise candidates" or stabilizing choices in times of institutional ambiguity \citep{burt2010neighbor}.

Third, the capacity for coalition formation often hinges on the interplay two network properties \citep{reagans2003network}: deep integration within cohesive subgroups and simultaneous reach across the broader structure of the College of Cardinals. This mechanism reflects the interplay between bonding and bridging social capital \citep{adler2002social}. Candidates embedded in dense, tightly-knit networks benefit from bonding capital, which provides trust, mutual obligation, and coordinated action. These close-knit groups function as electoral base camps, where early support can be consolidated, communication is rapid and reliable, and strategic coordination is facilitated \citep{Coleman1988}. Such structure plays a crucial role in the early rounds of voting, where uncertainty is high and the field of contenders still fluid. The internal solidarity of these subgroups enables them to act as cohesive voting blocs, amplifying their influence by signaling credibility and organizational capacity to the rest of the electorate. In this way, candidates from within these groups gain an initial momentum advantage, which can be essential in shaping perceptions of viability and setting the tone for broader coalition-building. At the same time, the ability to extend relational ties beyond the subgroup—that is, to engage across factional or regional divides—is what transforms a local base into a platform for wide-reaching support. Whereas isolated factions may foster loyalty, their insularity limits the candidate’s capacity to scale influence. By contrast, candidates who can anchor themselves in cohesive communities while also maintaining outward-facing ties are uniquely positioned to orchestrate cross-cutting alliances \citep{burt2005brokerage,obstfeld2005social}. Their social position enables them to translate local solidarity into broader legitimacy, enhancing their appeal as coalition candidates capable of unifying diverse constituencies within the College. In sum, this dual network configuration strengthens not only a candidate’s organizational support base but also their symbolic capacity to unify, making them especially effective in navigating the political complexity of conclave dynamics.

In theory, the candidate who succeeds may benefit from a combination of status, brokerage, and coalition-based advantages. Specifically, a cardinal who is simultaneously connected to prominent others, positioned between distinct peers, and able to activate ties with other communities may accumulate support organically over successive ballots. In this way, the conclave’s apparent unpredictability masks a deeper process of relational convergence, driven by the structure of the electoral network itself.

\section{Data and Methods}
\subsection{Data Collection}
To investigate the network mechanisms that permeate the conclave, we collected a broad range of data on current members of the College of Cardinals from publicly available sources. Data collection was conducted following two different logics. The first approach was informed by the assumption that the social structure connecting cardinals possesses a dual nature: it reflects both the current organizational configuration and historically constructed relationships whose influence persists over time. In addition to that, participation in collegial bodies cannot be equated with membership in a peer group, because the various Roman Curia entities that we investigated include hierarchical roles, such as secretary and under-secretary. Thus, we conducted an initial study focusing exclusively on the relationships among cardinals occupying top positions, while adopting a historical perspective that could also capture past co-membership ties among them within these collegial bodies. To construct the overall network, this affiliation network was subsequently overlaid with another network derived from consecration relationships. Additionally, we also included informal relationships among cardinals, which were identified through reputable journalistic articles.\footnote{https://www.catholicculture.org; https://www.pillarcatholic.com; https://www.indcatholicnews.com; https://angelusnews.com/.}  As mentioned, findings from this first study were disseminated to a broad audience via social media and newspaper publications before the election of Cardinal Prevost as Pope Leo XIV. For completeness, results and graphs from this data collection are fully reported in the Appendix of this article. It is important to note that the results of this first analyses and those of the more comprehensive investigation presented in this article are generally convergent in identifying key network figures.

The second approach and study, which is the focus of this paper, involved the reconstruction of a broader social structure encompassing all members of the collegial bodies, not just top roles, as of the date of the data collection period (April 2025). Specifically, we collected data through official Vatican archives, which we used to obtain information regarding the composition of the collegial bodies of the Roman Curia (e.g., commissions, dicasteries, etc...), and thus the institutional affiliations of all cardinals involved with these units.\footnote{https://press.vatican.va/} Furthermore, to ensure the accuracy of the data regarding the membership composition, we checked the affiliation of each cardinal by consulting their own biographical profiles,\footnote{https://press.vatican.va/content/salastampa/en/documentation/card\_bio\_typed.html}  which record their active memberships in the curial bodies. We did so for both voting and non-voting cardinals. Despite these careful checks, however, it is possible that a small number of ties may be missing due to official sources not being fully up to date or listing affiliations across multiple, inconsistently structured sections. However, we have reasons to believe that such instances are likely to represent a minority of cases. Moreover, prior research in social network analysis has shown that centrality measures tend to be robust to small amounts of missing data \citep{borgatti2006robustness}. As such, we do not expect these potential omissions to meaningfully affect our results. For the purposes of this study, we limited our analysis to active organizational affiliations in the collegial bodies as of the extraction date (29 April 2025). By anchoring our dataset to a specific date, we aimed to construct a balanced and consistent representation of the curial landscape at the time of the conclave. Such a standardizing of the temporal frame and sourcing methods also facilitates meaningful comparisons with future research on the College of Cardinals.

Following the same logic and data collection approach used previously, the network connecting the cardinals is constructed using a multiplex approach, that is, by overlaying the organizational co-memberships network with that of episcopal lineage obtained through the identification of living cardinals who are connected to one another through episcopal consecration ties. This information was obtained by consulting the Catholic-Hierarchy website, where each cardinal’s individual page provides details on their episcopal consecrator and co-consecrators.\footnote{https://www.catholic-hierarchy.org/bishop/scardc3.html}

Finally, we also collected additional information on each cardinal’s age (as of April 2025) and nationality using the same website. To support our visualizations and better contextualize the network, we incorporated data on each voting cardinal’s doctrinal orientation regarding key pastoral issues. These include positions on topics such as the blessing of same-sex couples, the ordination of female deacons, communion for divorced and remarried Catholics, and the decentralization of Church authority. These stances, which can be drawn from public statements, writings, and appointments, help illustrate underlying doctrinal alignments within the College of Cardinals. In our case, this information was obtained from an analysis published by \textit{The Sunday Times}, which classified cardinals based on their public statements along a spectrum ranging from strong liberal to strong conservative.\footnote{https://www.thetimes.com/uk/religion/article/inside-papal-conclave-politics-wjb9dhdvs}$^{,}$\footnote{All data we described is accessible at the following link: https://github.com/leonardorizzo/Conclave2025}

\subsection{Data Processing}
We begin our analysis by constructing a two-mode (i.e., bipartite) network that captures all affiliations between cardinals and the various collegial bodies of the Roman Curia, allowing for a clear visualization of institutional co-memberships. Figure \ref{fig:twomode} visualizes this network, which includes 160 nodes (136 cardinals and 24 entities) and 352 edges.

\begin{center}
\underline{INSERT FIGURE \ref{fig:twomode} ABOUT HERE}
\end{center}

Given the bipartite structure of the affiliation network, we performed a one-mode projection onto the set of cardinals. In this projected network, an undirected edge is established between any two cardinals who share membership in the same institutional body. Effectively, this means that each dicastery or affiliated entity is represented as a complete subgraph (clique) among its members. The resulting projection includes 136 cardinals and a total of 2792 weighted edges. Figure \ref{fig:formal} shows the projected network, highlighting the core–periphery structure that emerges from the data. In the visualization, the size of a node is scaled to reflect each cardinal’s weighted degree (i.e. the number of formal institutional connections they hold within the projected network), while the node color represents the cardinal’s doctrinal orientation—from liberal to conservative.

\begin{center}
\underline{INSERT FIGURE \ref{fig:formal} ABOUT HERE}
\end{center}

The second step involved constructing the co-consecration network. Although the consecration data naturally lends itself to a directed network representation, we opted to model it as an undirected network. This choice aligns with our analytical focus on identifying reciprocal ties between cardinals, without embedding directional assumptions into the structure. The resulting network, shown in Figure \ref{fig:cons}, comprises 123 nodes and 106 edges, and is organized into 24 distinct connected components.

\begin{center}
\underline{INSERT FIGURE \ref{fig:cons} ABOUT HERE}
\end{center}

We then integrated the formal co-membership network with the co-consecration network, resulting in a multiplex network comprising 188 cardinals and 2,884 weighted edges. Table \ref{tab:nets-descr} summarizes the key network statistics and associated data sources. Figures \ref{fig:multiplex} and \ref{fig:graph_tool} shows the resulting networks. In particular, while Figure \ref{fig:multiplex} leverages a standard spring-embedding layout algorithm, Figure \ref{fig:graph_tool} presents a community-based visualization of the network using edge bundling, which improves readability by highlighting intra-community cohesion and inter-community linkages \citep{holten2006hierarchical}.

\begin{center}
\underline{INSERT TABLE \ref{tab:nets-descr} ABOUT HERE}

\underline{INSERT FIGURE \ref{fig:multiplex} AND \ref{fig:graph_tool} ABOUT HERE}
\end{center}

\section{Network Analysis}

To identify the structural foundations of potential influence within the conclave, we analyzed the network positions occupied by cardinals in the College of Cardinals. As mentioned, the network we analyzed embodies both co-membership ties, which arise from joint service in collegial bodies, and consecration ties, reflecting apostolic lineage. We examined the three mechanisms introduced earlier—status, mediation power, and coalition building—and operationalized each of them using well-established network indicators.\footnote{Centrality measures were computed on the unweighted network. Robustness checks using weighted ties yielded consistent results (Section \ref{sec:robustness}).} For status, we used eigenvector centrality to capture the extent to which a cardinal is well connected to other central cardinals \citep{podolny2005status}. For mediation power, we computed betweenness centrality as a proxy for a cardinal’s capacity to bridge across disconnected parts of the network \citep{freeman1977set}. For coalition building, we computed a composite index that integrates three key parameters: community density, the size of the egonetwork, and the inter-community reach.

As mentioned, even if our analysis was released before the outcome of the conclave was announced, this is a descriptive case study, not predictive a model, and it aims to illuminate how different dimensions of structural advantage may pinpoint who is a strong \textit{papabile}.\footnote{We excluded from the rankings those cardinals who were not admitted to the conclave due to their age (above 80 years old). In theory, there is no formal age restriction for becoming Pope; however, with very few exceptions, only participants in the conclave have been elected to the papacy.}  Therefore, since our goal is to describe structural prominence of the members of the conclave, we do not adjust scores based on the cardinals’ age. Age is used solely to distinguish between those eligible to participate in the conclave and those who are not; all reported rankings refer exclusively to the former. Below, we present the empirical patterns associated with each of these three mechanisms.

\subsection{Status}
Eigenvector centrality captures not only how many connections an actor has, but also how well-connected their neighbors are, thus making it especially suited for identifying those who are embedded within the core of the network’s power structure \citep{bonacich2007some,aslarus2025early}. In this context, a high eigenvector score suggests that a cardinal is not only prominent, but connected to other prominent actors, and thus enjoys high structural visibility. The ranking in Table \ref{tab:eig} below shows the top 15 cardinals by eigenvector centrality. These individuals occupy structurally privileged positions that may afford them greater informal influence, visibility, and endorsement potential within the conclave.

\begin{center}
\underline{INSERT TABLE \ref{tab:eig} ABOUT HERE}
\end{center}

As the table shows, Cardinals Prevost, Tagle, and Parolin, who at the time of the conclave held the key formal role of Secretary of State, emerge as the most centrally embedded actors in terms of eigenvector centrality, followed closely by Cardinals Gugerotti and Fernández. Their high centrality scores reflect not only extensive ties within the College, but also the strategic quality of those ties, positioning them at the core of ecclesiastical network. These cardinals may be especially well-placed to attract early attention, mobilize endorsements, or signal institutional continuity within the deliberative dynamics of the conclave.

\subsection{Mediation Power}

We calculated each cardinal’s betweenness centrality on the combined network of co-membership and consecration ties. Betweenness centrality measures how frequently a node lies on the shortest paths between all other pairs of nodes \citep{freeman1977set}, capturing an actor’s capacity to connect disparate parts of the network. Table \ref{tab:betw} presents the top cardinals ranked by betweenness centrality:

\begin{center}
\underline{INSERT TABLE \ref{tab:betw} ABOUT HERE}
\end{center}

Cardinals Tagle, Prevost, and Betori emerge as powerful brokers. Figures like Cardinal Fernández and Cardinal Tobin also appear prominently, reflecting their role in connecting across different communities within the College of Cardinals. Interestingly, some cardinals such as Prevost, Tagle, Parolin, and Fernández who ranked highly on eigenvector centrality (status), also appear here—suggesting a dual advantage: they are both visible within central clusters and capable of brokering across subgroups. This convergence of structural prominence and brokerage potential may afford them particular influence in informal consensus-building processes, even if not all high-betweenness cardinals are considered \textit{papabili} in public discourse.

Some cardinals outside the conclave exhibit notable levels of mediation power, a result primarily driven by the inclusion of the consecration network. As it is possible to see in Figure \ref{fig:cons}, cardinals such as Re and Bertone have co-consecrated a significant number of current cardinals, thereby occupying structurally influential positions within the broader network despite being ineligible to vote.

\subsection{Coalition Building}
We developed a composite index that captures the structural features enabling a cardinal to activate and coordinate support across the broader network. As mentioned, this index combines three key components: community cohesion, egonetwork size, and cross-community reach. It is designed to reflect both a cardinal’s prominence within their immediate group and their ability to connect across group boundaries—an essential quality for forming broad alliances in a fragmented field of electors.

The first step involves the identification of communities within the network. Community detection seeks to locate densely connected regions—subsets of nodes that are more strongly interconnected relative to the rest of the network—and assign them to the same group. While numerous algorithms exist for this purpose, many are stochastic in nature and require the calibration of one or more parameters. To ensure reproducibility and reduce methodological complexity, we employed a deterministic approach based on greedy modularity optimization \citep{clauset2004finding}.\footnote{The algorithm identified 10 communities within the connected component, with a modularity score of 0.19.} It is important to note that the metrics presented in the subsequent analysis are highly sensitive to the underlying community structure, and may vary substantially depending on the chosen detection method.

The Coalition Building Index is calculated as a weighted average of the three components, each assigned a specific weight. Community cohesion is defined as the internal density of the cardinal’s community, expressed as a value between 0 and 1. Egonetwork size is measured by the cardinal’s number of direct connections (degree), normalized by the maximum degree in the network. Cross-community reach captures the share of other communities to which a cardinal is directly connected; this value also ranges between 0 and 1. Together, these elements provide a replicable and theory-informed measure of a cardinal’s coalition-building capacity based solely on their network position. Table \ref{tab:coal} below presents the top cardinals by coalition-building index:

\begin{center}
\underline{INSERT TABLE \ref{tab:coal} ABOUT HERE}
\end{center}

Cardinals Tagle, Czerny, and Parolin top the ranking, reflecting a high capacity to reach structurally diverse coalitions. Their profiles combine broad reach across communities, large egonetworks, and prominent positions within key subgroups. Notably, several of these cardinals also appeared in the top ranks of eigenvector and betweenness centrality, indicating once again potential convergence across mechanisms. Figures like Mamberti, Burke, Koch, and Prevost are also notable for their ability to span multiple groupings, which could be particularly relevant in a conclave characterized by fragmented doctrinal and regional blocs. This composite index thus highlights not only the potential for influence but also the relational infrastructure needed to anchor a viable candidacy, should momentum begin to form around these individuals.

\subsection{Underdogs and Kingmakers: The Network Advantage of Cardinal \mbox{Prevost}}
When it comes to forecasting frontrunners, many tend to favor visible markers of influence: tenure, title, charisma. The \textit{papabili} lists that circulated ahead of the 2025 conclave were no exception. With very few exceptions, pundits, commentators, and bookmakers clustered around highly visible members of the Roman Curia and prominent archbishops, that is, those with formal authority, diplomatic profiles, or public recognition

Our analyses highlighted a different angle. Influential individuals in a closed system like the conclave are not only those with the biggest titles, but also those who occupy advantageous positions within the relational infrastructure of the Church, by virtue of being connected to prominent others or sitting at the crossroads of different communities. This is where network theory comes in—and where Cardinal Prevost stood out. As summarized in Table 4, he was not top of mind in media speculation and the public discourse, having a mere 1\% probability according to many bookmakers. He was a relative newcomer, with only two years in the College of Cardinals. Yet, in our network analysis, he quietly emerged as one of the most structurally advantaged actors in the College. He ranked highest in eigenvector centrality, and near the top in both and coalition building. In other words, he occupied the kind of position that makes people matter: at the center of consensus and between otherwise disconnected factions.

\begin{center}
\underline{INSERT TABLE \ref{table-bet} ABOUT HERE}
\end{center}

This contrast is striking. While public forecasts emphasized surface-level attributes, our structural indicators captured less visible but meaningful markers of influence. Prevost was not the most prominent figure publicly, yet he demonstrated a high likelihood of connecting the right individuals at critical moments. The \textit{papabili} lists, grounded in visible credentials, overlooked insights revealed through the network. The election of Cardinal Prevost—Pope Leo XIV—was not just the result of personal qualities. It was also the result of a strong structural position.

Our model also highlights another similar case to Cardinal Prevost: Cardinal José Tolentino de Mendonça. Despite occupying a structurally prominent position within the network in some structural dimensions, he was largely absent from public \textit{papabili} lists. One likely reason is his age: at 59, he may have been perceived as relatively young to be considered a serious contender. Yet it is worth noting that he is not significantly younger than Cardinal Pizzaballa, who appeared in many predictive models. This suggests that Cardinal Tolentino’s low visibility was less a matter of eligibility and more a reflection of how traditional analyses continue to overlook structural indicators of influence.

\subsection{Robustness Analyses}\label{sec:robustness}
In this network analysis, we excluded \textit{prefects emeritus} from the institutional membership data, as these positions are largely honorary and typically held by cardinals over the age of 80—thus outside the electorate of the conclave.  To test the robustness of our findings, we reran the analysis including these emeritus affiliations, which added approximately 300 additional edges to the network. The overall results remained stable, indicating that these roles do not significantly affect the positioning of the main contenders within the network.

Additionally, we tested the impact of incorporating the informal ties previously identified through journalistic sources. Due to their limited number, these ties had no significant effect on the overall network structure or rankings. Likewise, substituting weighted variants of centrality measures—accounting for the strength of ties—for eigenvector and betweenness centrality did not yield substantial changes in the cardinal rankings.

We also examined both weighted and unweighted degree centrality as alternative metrics and found consistent results: Cardinal Prevost ranked first, further reinforcing his prominence across different dimensions of network influence. In addition, we computed a measure of 2-step local eigenvector centrality to address a potential limitation of standard eigenvector centrality, which can assign disproportionately high scores to nodes embedded in large cliques. By focusing on influence within a two-step neighborhood, this measure reduces the inflationary effect of cliquishness. Even with this approach Cardinal Prevost remained at the top of the status ranking.

Finally, we conducted a focused analysis using only the formal institutional network, excluding all consecration ties. Under this configuration, Cardinal Prevost’s network centrality increased significantly across multiple dimensions. He ranked first across all of the three dimensions, underscoring his dominant structural position within the formal curial network. This result highlights Prevost’s strategic advantage in the realm of institutional affiliations, even in the absence of ecclesiastical lineage ties.

All of these results are reported as raw values, that is, without applying the age adjustment (see Appendix \ref{sec:age-adj}). Importantly, the findings remain consistent when the age-adjustment factor is implemented, indicating the robustness of the cardinal rankings across both adjusted and unadjusted measures.

\section{Discussion}
Our reconstruction of the College of Cardinals as a multiplex network offers a powerful lens through which to explore the complex dynamics of the conclave: existing alongside the formal hierarchical structure, this network provides cardinals with both opportunities and constraints in mobilizing critical resources consequential to the decision-making process. The conclave may be wrapped in ritual, but influence still follows relational pathways that are as patterned, and as measurable, as those in any corporate boardroom or parliamentary council \citep{westphal1997defections}. By foregrounding structure, we reaffirm a foundational insight of network research: position is a structural signature of influence.

As we discussed, the practical payoff of this lens is summarized by how Cardinal Prevost became Pope Leo XIV. On the eve of the conclave most bookmakers\footnote{Following a CNBC report, bettors poured more than \$40 million into the papal conclave, on just two prediction market platforms: over \$30 million on Polymarket, according to its website, and over \$10.6 million on Kalshi, says a company spokesperson. Cardinal Prevost had odds of less than 1\%, according to Kalshi’s website. Of the over 33,000 conclave trades placed on Kalshi, just 416 trades totaling to nearly \$450,000 were placed on Prevost.} and pundits relegated him to the margins, overshadowed  instead by more visible media favorites. The visibility of these others candidates was driven by individual attributes that were easily observable from the outside. For instance, Cardinal Parolin was considered a frontrunner due to his formal organizational rank as Secretary of State. Cardinal Tagle, by contrast, was recognized by public opinion for his extroverted style, participatory leadership, and attentiveness to youth issues. Cardinal Besungu, in addition to his extensive experience, was seen as the most plausible candidate from the African continent, which has never in its history produced a pope. Cardinal Dolan was viewed as papabile in the event that a conservative candidacy prevailed in the conclave, while Cardinal Zuppi was among the leading figures associated with the progressive camp. The network perspective we adopted offers a different angle to understand the outcome of such a complex decision-making process. Specifically, the roles played within the co-membership network, which are visible within the College of Cardinals, and the informational leverage derived from his structural position placed Prevost squarely among the set of credible frontrunners.

The tension between attributes and relationships in shaping social outcomes is particularly salient in the context of the papal conclave, where it becomes visible in the contrast between prevailing media narratives and our network analysis. Public commentary has largely focused on observable individual characteristics—such as age, nationality, theological orientation or the formal role within the Roman Church—as key indicators of papal viability. In contrast, our approach highlights structural embeddedness as a more powerful, albeit less visible to the outside, marker of influence. This divergence recalls the foundational argument of \citet{wellman1988social}, who emphasized how behavior is often better explained in terms of relations rather than attributes, while also acknowledging their interaction.

Our analysis echoes the idea that network position is a robust correlate of behaviors, including voting decision as in the conclave. Yet we do not claim that structural advantage exists independently of individual traits or agency. A cardinal known for his doctrinal and political moderation, for instance, may be more likely to span ideological divides, suggesting that centrism operates not only as a personal attribute but also as a facilitator of bridging ties. This entanglement complicates causal inference: does influence stem from theological stance, from structural position, or from the mutually reinforcing alignment between the two? In other words, the relationship between personal characteristics and structural embeddedness raises important questions of endogeneity. From a theoretical perspective, several mechanisms may plausibly connect cardinals’ attributes, their structural positions, and their likelihood of receiving votes. In an additive logic, attributes and network positions may have an independent and direct effect on outcomes \citep{mehra2001social}. Alternatively, attributes may shape network structure—self-monitoring, for example, may facilitate cross-cutting ties \citep{casciaro2015integration, sasovova2010network}—thus indirectly influencing vote outcomes through relationships. Conversely, a cardinal’s central position in the network may shape his leadership skills or increase the perceived legitimacy of his traits, making structure an antecedent of malleable traits and skills. Furthermore, individual attributes and social structures could interact to explain behaviors. This idea is in line with “interaction models” of network which suggest how outcomes are shaped by a complementarity effect between structural network positions and individual traits \citep{soda2018harvesting}.

In this discussion, however, it is important to consider the nature of ties which represents an important boundary condition. We study a context where ties among the cardinals emanate from a co-membership networks based on institutionalized patterns of shared committee service and episcopal lineages. These ties are formally mandated and thus relatively stable. Unlike informal interpersonal ties, these connections are not easily initiated or dissolved through individual agency, at least in the short term. As such, our networks are less plastic and less prone to ego-centric manipulation \citep{tasselli2021network}.

As the networks examined are the result of organizational design rather than the outcome of a self-organizing process, the role of agency and individual factors is attenuated or at least unfolding through actions aimed at influencing organizational decision-makers \citep{westphal2007flattery}. It was the organizational decisions regarding the composition of collegial bodies that determined Prevost’s structural prominence. In this respect, Cardinal Prevost’s career within is particularly revealing. His trajectory within the Roman Curia was notably accelerated under Pope Francis, who entrusted him with a series of increasingly influential roles. Beginning with his appointment as Apostolic Administrator of Chiclayo in 2014, Prevost was subsequently named bishop of the same diocese in 2015, and progressively took on responsibilities both in Peru and within the universal Church. His appointments to multiple key dicasteries—including the key Dicastery for Bishops and the Congregation for the Clergy—culminated in his 2023 nomination as Prefect of the Dicastery for Bishops (which has the important task of selecting and supervising bishops) and President of the Pontifical Commission for Latin America, all positions of substantial influence in the governance of the Church. Created cardinal in September 2023, he entered the order of cardinal bishops in February 2025, receiving the suburbicarian see of Albano. To what extent all these appointments, which enhanced Cardinal Prevost’s prominence and visibility, were the result of an intentional design by Pope Francis and part of a plausible “succession plan” is impossible to determine.

In sum, what we have uncovered it that in a highly complex organization, the composition of collegial bodies—much more than formal hierarchy—facilitates the formation of networks that are highly relevant to key decision-making processes and that networks within organizations tend to be draped around the scaffolding of formal arrangements \citep{soda2025organization}.

Contributing to organizational network research, our study highlights the relevance of social infrastructures designed by organizational choices that foster connections among members beyond hierarchical ties. In the case of the conclave, the relational space generated through co-participation in collegial bodies played an important role in shaping voting behavior and consensus-building processes. More broadly, our study underscores the importance of accurately reconstructing all relational spaces—both formal and informal—that emerge within organizations, in order to better understand their underlying dynamics.

From a network intervention standpoint, initiatives aimed at shaping interpersonal relationships and the structures of intra-organizational networks could be effectively anchored in design choices \citep{ouchi1974defining}. Although these interventions are initiated through formal changes—such as alterations to structures, processes, and organizational systems—their effects on social structures can be systematically analyzed using network analysis methodologies \citep{valente2012network}.

Another implication of our study is that it offers a cautionary note for the current infatuation with big data and the use of AI. Predictive dashboards and AI pipelines that scrape speeches, tweets, or betting markets missed Cardinal Prevost because they modelled attributes and signals, while forgetting relationships \citep{antonioni2025complex}. The implication is not that machine learning is futile, but that strategic value lies in asking the right questions and using the right theoretical lenses. In this study, we argued that social network analysis is a powerful tool to investigate the conclave. Needless to say, this approach is deeply rooted in a long tradition of studies and research. From \citeauthor{katz1957two}’s (1957) diffusion thresholds to \citeauthor{freeman1977set}’s (1977) betweenness to Granovetter (1973) weak ties, the literature on networks has long showed that position matters. The conclave reminds us that classical tools remain potent when applied with care. By projecting a bipartite affiliation structure and layering a sacramental lineage onto it, we have shown how simple techniques illuminate a domain that, at first glance, seems empirically intractable.

While our study sheds light on the structural underpinning associated with being \textit{papabile} within the College of Cardinals, several limitations deserve careful attention. First, our network reconstruction is based entirely on formal ties—co-membership in ecclesiastical bodies and consecration lineages. These ties offer a meaningful proxy of important relationships, but they inevitably miss the informal social ties among cardinals. Friendship ties, informal mentoring relationships, and personal rivalries or negative ties \citep{kilduff2024hiding,labianca2006negative}, which could play a key role in decision making, remain outside the scope of our investigation. Second, our arguments linking the prominence of Cardinal Prevost to his position in the Vatican network assume that co-membership and co-consecration ties are visible to the College of Cardinals, while opaque for external observers. Specifically, within the College of Cardinals, there is high visibility of the roles underlying co-membership ties and a clear awareness of consecration links. However, while visible, interpersonal ties may or may be not perceived accurately by cardinals. In this regard, we do not capture perceived cognitive social structures, meaning that we don’t know how cardinals perceive the network they are embedded in. Research has shown that actors often rely on subjective network maps when making strategic decisions \citep{krackhardt1987cognitive}, and these maps can diverge sharply from objective structure. A cardinal may appear central in our reconstruction, yet be perceived as peripheral by his peers, or vice versa \citep{iorio2022brokers}. Without access to first-hand perceptions, interviews, or internal documents, we cannot account for this cognitive dimension of influence \citep{aslarus2025early}. Third, our approach is cross-sectional and static. It does not model how the network evolves in the lead-up to the conclave or preferences shift across successive ballots. Nor do we account for external interventions (e.g., regional lobbying) that may reconfigure priorities in real time. Finally, our analysis is bounded by observability and survivorship: we analyze the College as it existed in April 2025, but the relational dynamics that matter most may have formed over decades of prior interaction, some of which leave no public trace. Future research may explore ways to approximate these hidden histories—through qualitative fieldwork, ecclesiastical archives, or digital records of public appearances.

Finally, while our analysis deliberately stops short of making predictions, it offers a structural map that narrows the range of plausible outcomes—and helps illuminate why some candidacies crystallize while others fade.

\section{Conclusion}
The broader takeaway of our study is straightforward: no serious analysis of future conclaves can afford to ignore network structure. Whether scholars pursue qualitative thick description, statistical modelling, or computational simulation, the relational scaffolding of the College must serve as the baseline upon which other factors—doctrinal, regional, and theological—operate. The conclave is not merely a gathering of isolated cardinal electors; it is an interaction network with a long memory and a short time horizon. To attempt to decipher its logic, we must put a relational perspective back at the center of the stage.
\clearpage
\pagebreak
{
\singlespacing
\bibliographystyle{apalike}
\bibliography{bibliography}
}
\pagebreak
\appendix
\section*{Tables and Figures}

\begin{table}[H]
\caption{Networks summary}
\label{tab:nets-descr}
\begin{center}
\footnotesize
\begin{tabular}{|c|c|c|c|c|}
\hline
\textbf{Layer} & \textbf{Type} &  \textbf{Nodes} & \textbf{Links} & \textbf{Source} \\
\hline
\begin{tabular}{c}Roman\\Curia\end{tabular} & 
\begin{tabular}{c}Two-\\Mode\end{tabular} & 
\begin{tabular}{c}136\\Cardinals\\+ 24\\Vatican\\bodies\end{tabular} & 
352 & 
\footnotesize{https://press.vatican.va/} \\
\hline
\begin{tabular}{c}Roman\\Curia\\(projected)\end{tabular} & 
\begin{tabular}{c}One-\\Mode\end{tabular} & 
136 & 
2792 & 
\footnotesize{https://press.vatican.va/} \\
\hline
Consecration & 
\begin{tabular}{c}One-\\Mode\end{tabular} & 
123 & 
106 & 
\footnotesize{https://www.catholic-hierarchy.org/bishop/scardc3.html} \\
\hline
Multiplex & 
\begin{tabular}{c}One-\\Mode\end{tabular} & 
188 & 
2884 & 
-- \\
\hline
\end{tabular}
\end{center}
\end{table}

\begin{table}[H]
\caption{Top 15 Cardinals by Status}
\label{tab:eig}
\begin{centering}
\begin{tabular}{cc}
\tabularnewline
\textbf{Cardinal} & \textbf{Position within Conclave}\tabularnewline
\hline 
Prevost & 1\tabularnewline
Tagle & 2\tabularnewline
Parolin & 3\tabularnewline
Gugerotti & 4\tabularnewline
Fernández & 5\tabularnewline
Mendonça & 6\tabularnewline
You Heung-Sik & 7\tabularnewline
Roche & 8\tabularnewline
Sarah & 9\tabularnewline
Sturla Berhouet & 10\tabularnewline
Farrell & 11\tabularnewline
Turkson & 12\tabularnewline
Dew & 13\tabularnewline
Mamberti & 14\tabularnewline
Scherer & 15\tabularnewline
\end{tabular}
\par\end{centering}
\end{table}


\begin{table}[H]
\caption{Top 15 Cardinals by Mediation Power}
\label{tab:betw}
\begin{centering}
\begin{tabular}{cc}
\tabularnewline
\textbf{Cardinal} & \textbf{Position within Conclave}\tabularnewline
\hline 
Re & not in the conclave\tabularnewline
Bertone & not in the conclave\tabularnewline
Tagle & 1\tabularnewline
Prevost & 2\tabularnewline
Betori & 3\tabularnewline
Fernández & 4\tabularnewline
Romeo & not in the conclave\tabularnewline
Tobin & 5\tabularnewline
Parolin & 6\tabularnewline
Zuppi & 7\tabularnewline
Czerny & 8\tabularnewline
De Donatis & 9\tabularnewline
Semeraro & 10\tabularnewline
Koch & 11\tabularnewline
Calcagno & not in the conclave\tabularnewline
\end{tabular}
\par\end{centering}
\end{table}


\begin{table}[H]
\caption{Top 15 by Coalition building capacity}
\label{tab:coal}
\begin{centering}
\begin{tabular}{cc}
\tabularnewline
\textbf{Cardinal} & \textbf{Position within Conclave}\tabularnewline
\hline 
Tagle & 1\tabularnewline
Czerny & 2\tabularnewline
Parolin & 3\tabularnewline
Mamberti & 4\tabularnewline
Burke & 5\tabularnewline
Schönborn & not in the conclave\tabularnewline
Koch & 6\tabularnewline
Dolan & 7\tabularnewline
Prevost & 8\tabularnewline
Filoni & 9\tabularnewline
Erd\H{o} & 10\tabularnewline
Bertone & not in the conclave\tabularnewline
Ouellet & not in the conclave\tabularnewline
Baldisseri & not in the conclave\tabularnewline
Calcagno & not in the conclave\tabularnewline
\end{tabular}
\par\end{centering}
\end{table}

\begin{sidewaystable}
\begin{centering}
\caption{Cardinal Profiles\protect}
\label{table-bet}
\begin{tabular}{ccccccccc}
\tabularnewline
{\small\textbf{Name}} & {\small{}%
\begin{tabular}{c}
{\small\textbf{Tenure}}\tabularnewline
{\small\textbf{as}}\tabularnewline
{\small\textbf{cardinal}}\tabularnewline
\end{tabular}} & {\small\textbf{Age}} & {\small{}%
\begin{tabular}{c}
{\small\textbf{Formal}}\tabularnewline
{\small\textbf{rank}}\tabularnewline
\end{tabular}} & {\small{}%
\begin{tabular}{c}
{\small\textbf{Doctrinal}}\tabularnewline
{\small\textbf{Orientation}}\tabularnewline
\end{tabular}} & {\small\textbf{Status}} & {\small{}%
\begin{tabular}{c}
{\small\textbf{Mediation}}\tabularnewline
{\small\textbf{Power}}\tabularnewline
\end{tabular}} & {\small{}%
\begin{tabular}{c}
{\small\textbf{Coalition}}\tabularnewline
{\small\textbf{building}}\tabularnewline
{\small\textbf{capacity}}\tabularnewline
\end{tabular}} & {\small{}%
\begin{tabular}{c}
{\small\textbf{Polymarket}}\tabularnewline
{\small\textbf{probability*}}\tabularnewline
\end{tabular}}\tabularnewline
\hline 
 &  &  &  &  &  &  &  & \tabularnewline
{\small P. Parolin} & 11 & 70.2 & {\footnotesize{}%
\begin{tabular}{c}
{\footnotesize Secretary}\tabularnewline
{\footnotesize of State}\tabularnewline
\end{tabular}} & {\footnotesize Liberal} & 3 & 6 & 3 & 25.50\%\tabularnewline
{\small L. A. Tagle} & 13 & 67.8 & {\footnotesize{}%
\begin{tabular}{c}
{\footnotesize Pro-Prefect }\tabularnewline
{\footnotesize of the Dicastery}\tabularnewline
{\footnotesize for Evangelization}\tabularnewline
\end{tabular}}{\footnotesize{} } & {\footnotesize{}%
\begin{tabular}{c}
{\footnotesize Soft}\tabularnewline
{\footnotesize Liberal}\tabularnewline
\end{tabular}}{\footnotesize{} } & 2 & 1 & 1 & 19.65\%\tabularnewline
{\small M. Zuppi} & 6 & 69.5 & {\footnotesize{}%
\begin{tabular}{c}
{\footnotesize Archbishop}\tabularnewline
{\footnotesize of Bologna}\tabularnewline
\end{tabular}} & {\footnotesize{}%
\begin{tabular}{c}
{\footnotesize Soft}\tabularnewline
{\footnotesize Liberal}\tabularnewline
\end{tabular}}{\footnotesize{} } & 19 & 7 & 25 & 10.15\%\tabularnewline
{\small P. Pizzaballa} & 5 & 60.0 & {\footnotesize{}%
\begin{tabular}{c}
{\footnotesize Latin Patriarch}\tabularnewline
{\footnotesize of Jerusalem}\tabularnewline
\end{tabular}} & {\footnotesize{}%
\begin{tabular}{c}
{\footnotesize Soft}\tabularnewline
{\footnotesize Conservative}\tabularnewline
\end{tabular}} & 62 & 83 & 73 & 8.80\%\tabularnewline
{\small P. Turkson} & 22 & 76.5 & {\footnotesize{}%
\begin{tabular}{c}
{\footnotesize Chancellor of the}\tabularnewline
{\footnotesize Pontifical Academy}\tabularnewline
{\footnotesize of Sciences}\tabularnewline
\end{tabular}}{\footnotesize{} } & {\footnotesize Moderate} & 12 & 53 & 24 & 7.35\%\tabularnewline
{\small P. Erd\H{o}} & 22 & 72.8 & {\footnotesize{}%
\begin{tabular}{c}
{\footnotesize Metropolitan Archbishop}\tabularnewline
{\footnotesize of }\tabularnewline
{\footnotesize Esztergom-Budapest}\tabularnewline
\end{tabular}}{\footnotesize{} } & {\footnotesize{}%
\begin{tabular}{c}
{\footnotesize Soft}\tabularnewline
{\footnotesize Conservative}\tabularnewline
\end{tabular}} & 26 & 12 & 10 & 6.55\%\tabularnewline
{\small J.M. Aveline} & 3 & 66.3 & {\footnotesize{}%
\begin{tabular}{c}
{\footnotesize Metropolitan Archbishop}\tabularnewline
{\footnotesize of Marseille}\tabularnewline
\end{tabular}} & {\footnotesize{}%
\begin{tabular}{c}
{\footnotesize Soft}\tabularnewline
{\footnotesize Liberal}\tabularnewline
\end{tabular}}{\footnotesize{} } & 60 & 65 & 71 & 4.35\%\tabularnewline
{\small R. Sarah} & 15 & 79.8 & {\footnotesize{}%
\begin{tabular}{c}
{\footnotesize Prefect emeritus}\tabularnewline
{\footnotesize of the Dicastery}\tabularnewline
{\footnotesize of Divine Worship}\tabularnewline
\end{tabular}}{\footnotesize{} } & {\footnotesize Conservative} & 9 & 29 & 18 & 3.20\%\tabularnewline
{\small M. Grech} & 5 & 68.1 & {\footnotesize{}%
\begin{tabular}{c}
{\footnotesize Secretary General}\tabularnewline
{\footnotesize of General Secretary}\tabularnewline
{\footnotesize of the Synod}\tabularnewline
\end{tabular}}{\footnotesize{} } & {\footnotesize{}%
\begin{tabular}{c}
{\footnotesize Soft}\tabularnewline
{\footnotesize Liberal}\tabularnewline
\end{tabular}}{\footnotesize{} } & 40 & 52 & 53 & 2.30\%\tabularnewline
{\small F. A. Besungu} & 6 & 65.2 & {\footnotesize{}%
\begin{tabular}{c}
{\footnotesize Archbishop}\tabularnewline
{\footnotesize of Kinshasa}\tabularnewline
\end{tabular}} & {\footnotesize{}%
\begin{tabular}{c}
{\footnotesize Soft}\tabularnewline
{\footnotesize Conservative}\tabularnewline
\end{tabular}}{\footnotesize{} } & 81 & 86 & 77 & 1.45\%\tabularnewline
{\small J. T. de Mendonça} & 6 & 59.3 & {\footnotesize{}%
\begin{tabular}{c}
{\footnotesize Prefect of the}\tabularnewline
{\footnotesize Dicastery for}\tabularnewline
{\footnotesize Culture and Education}\tabularnewline
\end{tabular}}{\footnotesize{} } & {\footnotesize{}%
\begin{tabular}{c}
{\footnotesize Soft}\tabularnewline
{\footnotesize Liberal}\tabularnewline
\end{tabular}}{\footnotesize{} } & 6 & 17 & 11 & 1.40\%\tabularnewline
{\small R. F. Prevost} & 2 & 69.6 & {\footnotesize{}%
\begin{tabular}{c}
{\footnotesize Prefect of the}\tabularnewline
{\footnotesize Dicastery for}\tabularnewline
{\footnotesize Bishops}\tabularnewline
\end{tabular}} & {\footnotesize Moderate} & 1 & 2 & 8 & 1.15\%\tabularnewline
\end{tabular}
\par
*\footnotesize{Polymarket is a decentralized prediction platform where users trade on the outcomes of real-world events. Polymarket probabilities are the market-implied likelihoods of those outcomes, reflected in the price of each share.}
\end{centering}
\end{sidewaystable}


\begin{figure}[H]
\begin{centering}
\includegraphics[width=1\textwidth]{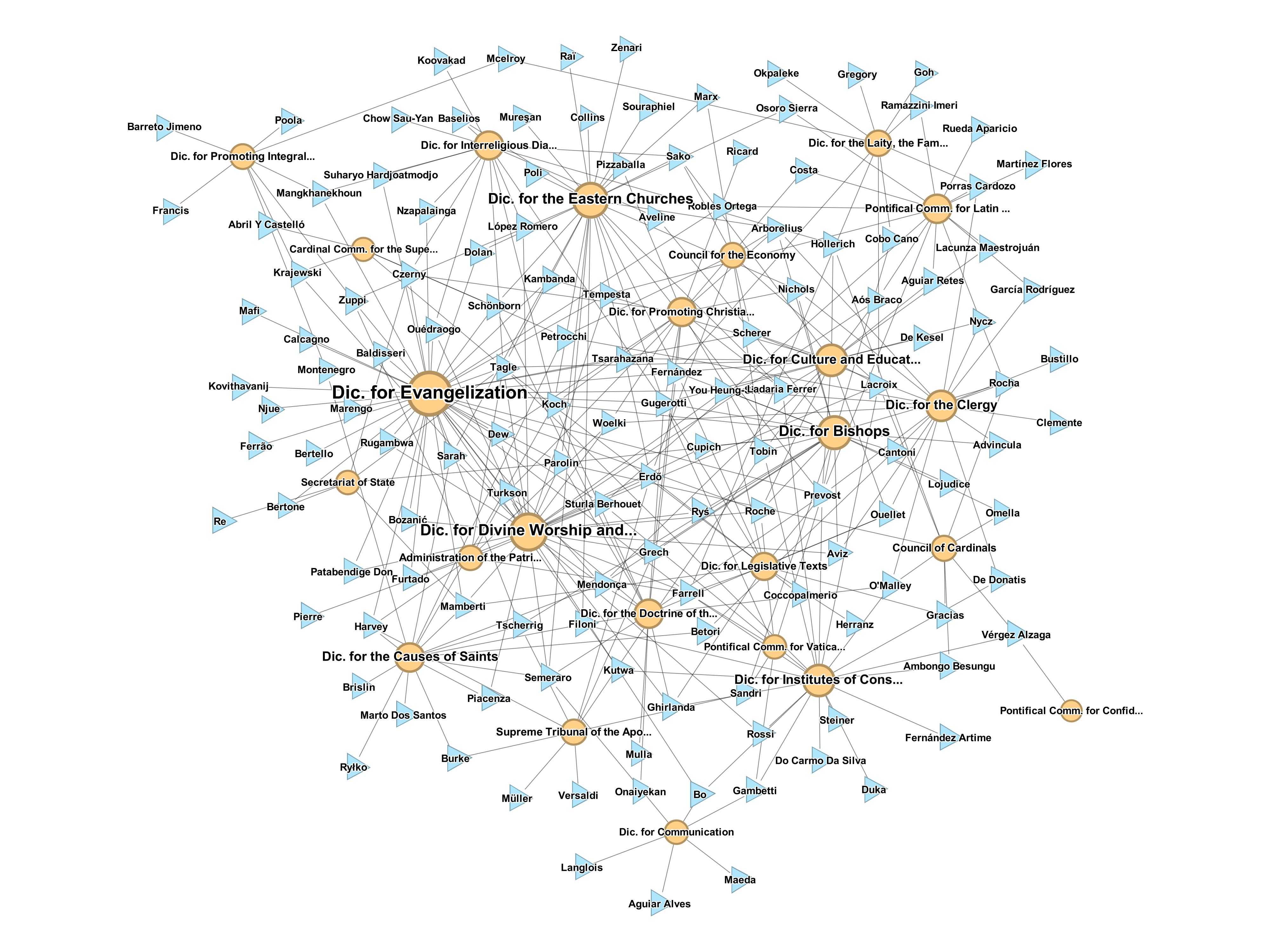}
\par\end{centering}
\centering{}\caption{Two-Mode Network of Collegial Bodies and Affiliated Cardinals.}
Collegial bodies are shown in orange and cardinals in blue. Cardinal nodes are uniform in size, while the size of collegial body nodes is proportional to the number of affiliated cardinals.

\label{fig:twomode}
\end{figure}

\begin{figure}[H]
\begin{centering}
\includegraphics[width=1\textwidth]{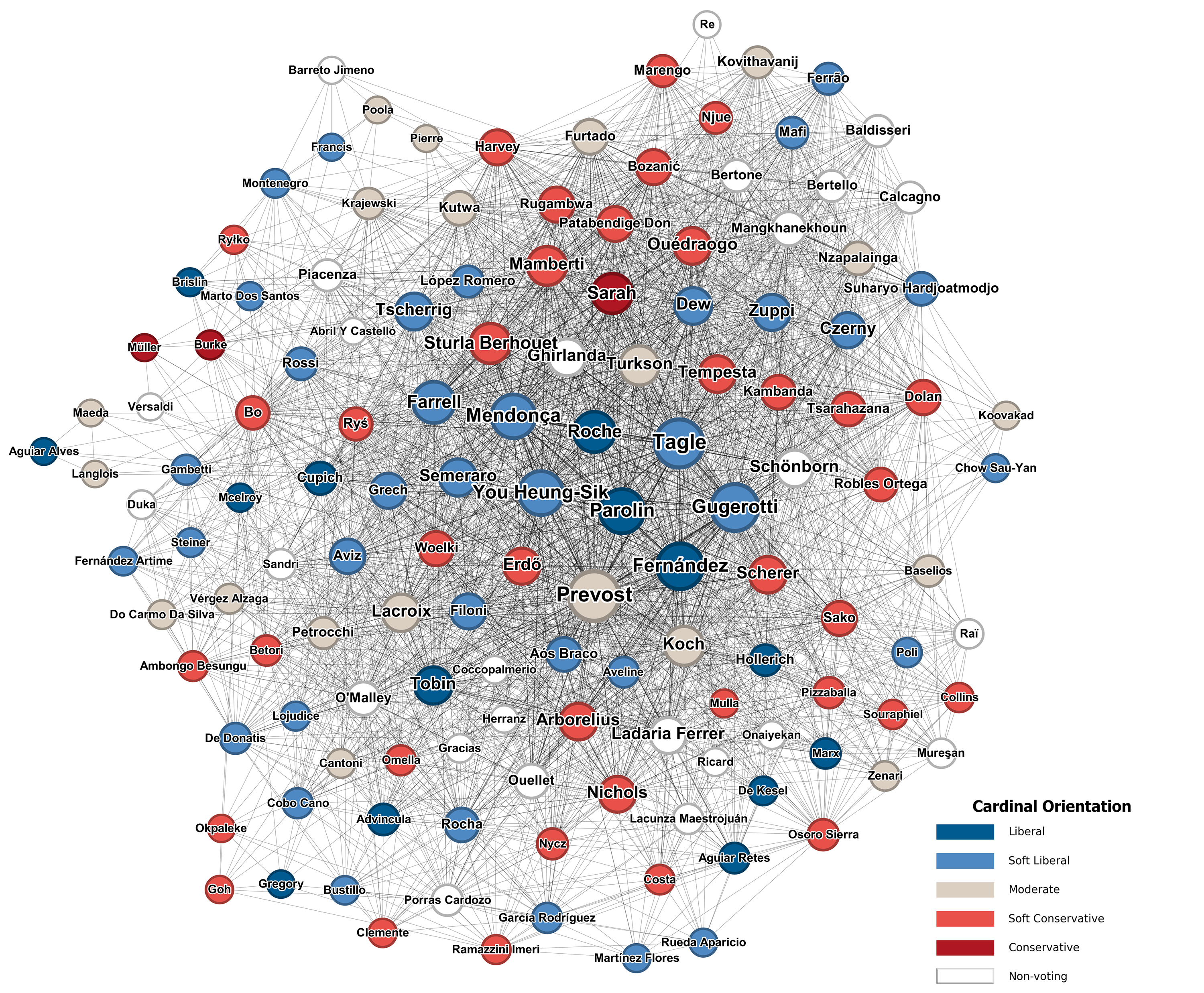}
\par\end{centering}
\centering{}\caption{Co-Membership Network of Cardinals.}
\label{fig:formal}
\end{figure}

\begin{figure}[H]
\begin{centering}
\includegraphics[width=1\textwidth]{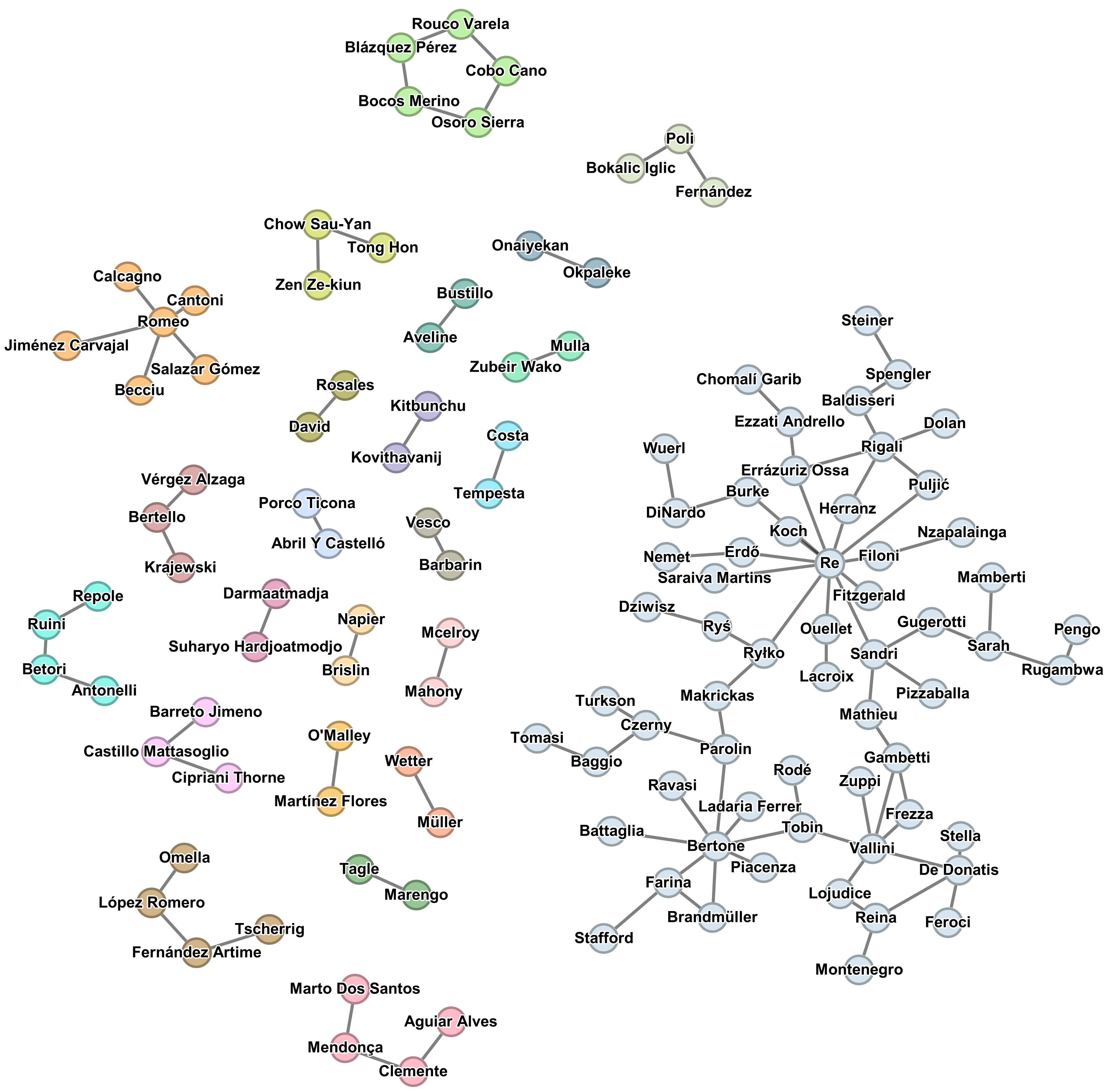}
\par\end{centering}
\centering{}\caption{Co-Consecration Network among Cardinals.}
\label{fig:cons}
Different colors are used for the 24 distinct connected components.
\end{figure}

\begin{figure}[H]
\begin{centering}
\includegraphics[width=1\textwidth]{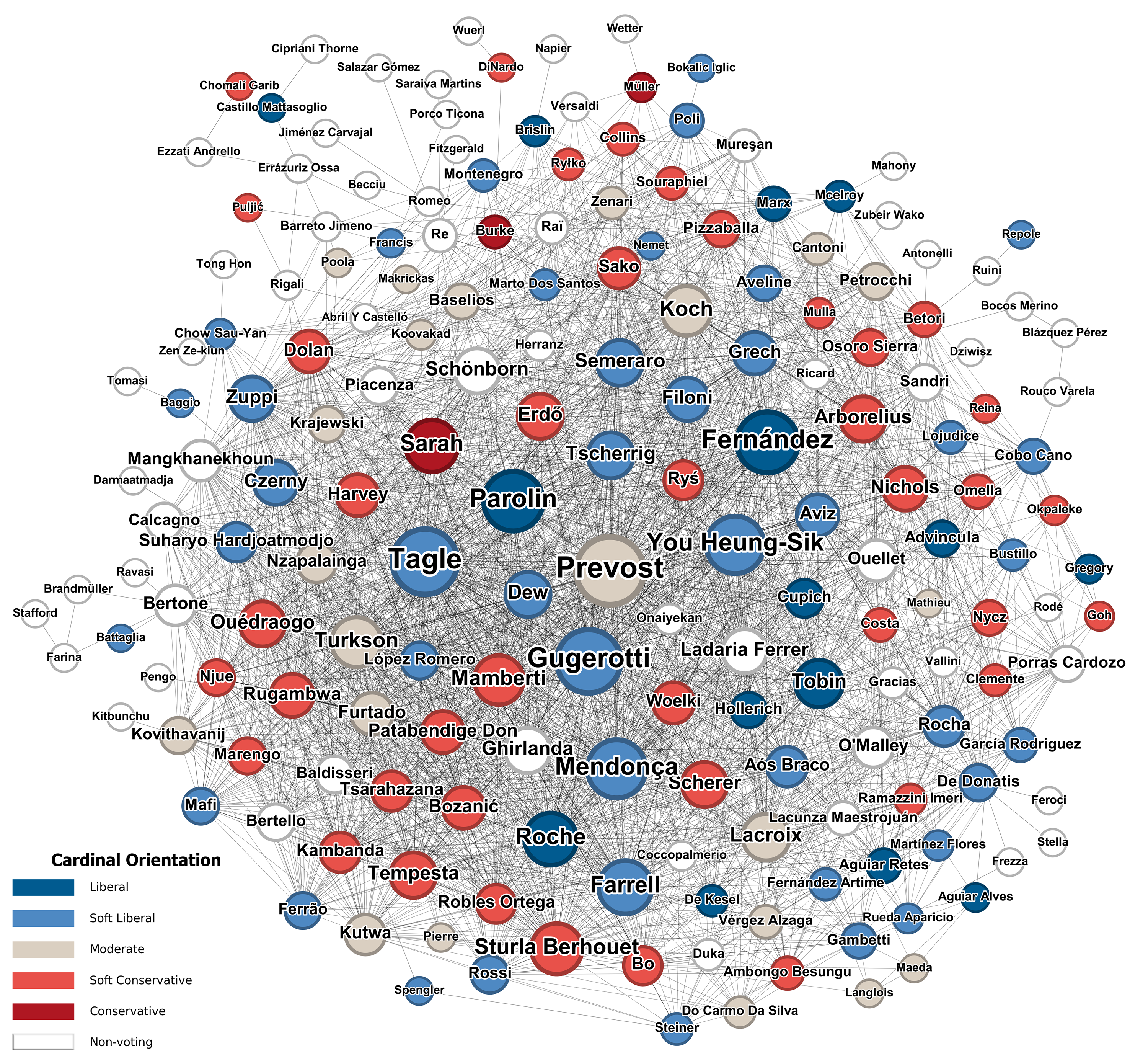}
\par\end{centering}
\centering{}\caption{Multiplex College of Cardinals Network Obtained from the Union of Co-Membership and Co-Consecration Ties.}\label{fig:multiplex}
\end{figure}

\begin{figure}[H]
\begin{centering}
\includegraphics[width=1\textwidth]{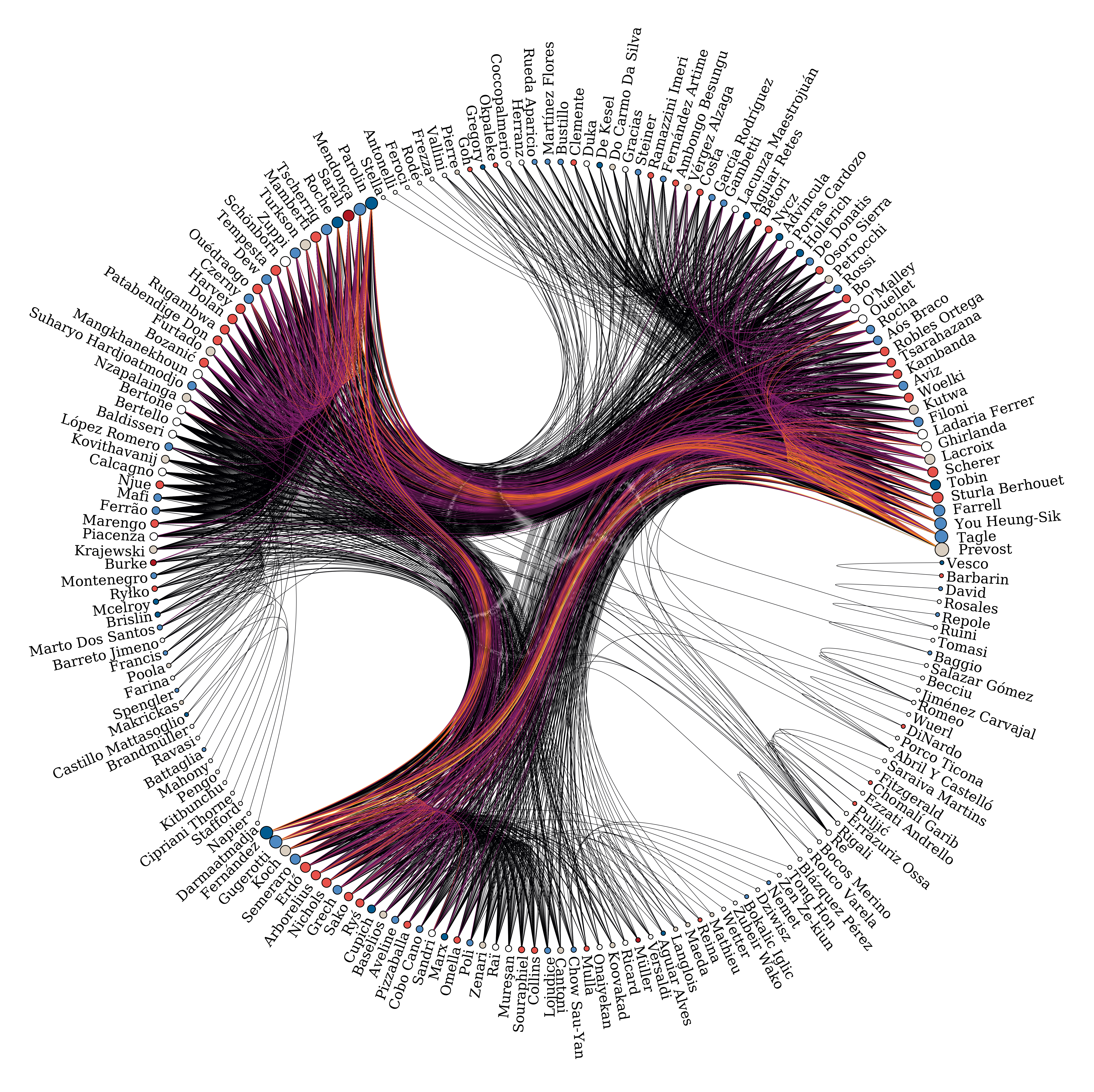}
\par\end{centering}
\caption{Multiplex Network of Cardinals Visualized as Community Structures.}
\label{fig:graph_tool}
Nodes are colored based on doctrinal view and sized based on eigenvector centrality. The figure shows that there are many more ties within communities than across them: Cardinals tend to collaborate with each other within communities more often than across them. But some cardinals do broker across communities. For example, the orange lines illustrate the relationships maintained by hubs Prevost, Tagle, Fernandez, Gugerotti, and Parolin who collaborate with members of the other communities.
\end{figure}

\pagebreak

\appendix
\renewcommand{\thefigure}{A\arabic{figure}}
\setcounter{figure}{0}  
\renewcommand{\thetable}{A\arabic{table}}
\setcounter{table}{0}
\section{Appendix: Online Article}

This appendix provides detailed information and results from the network analysis previously published online and in newspapers.

\paragraph{Data Collection}

Here, we report the findings obtained using the first logic of data collection, which was used in the analyses published by Corriere della Sera and Bocconi; see Figure A1 below.

\begin{figure}[h]
\begin{centering}
\includegraphics[width=0.7\textwidth]{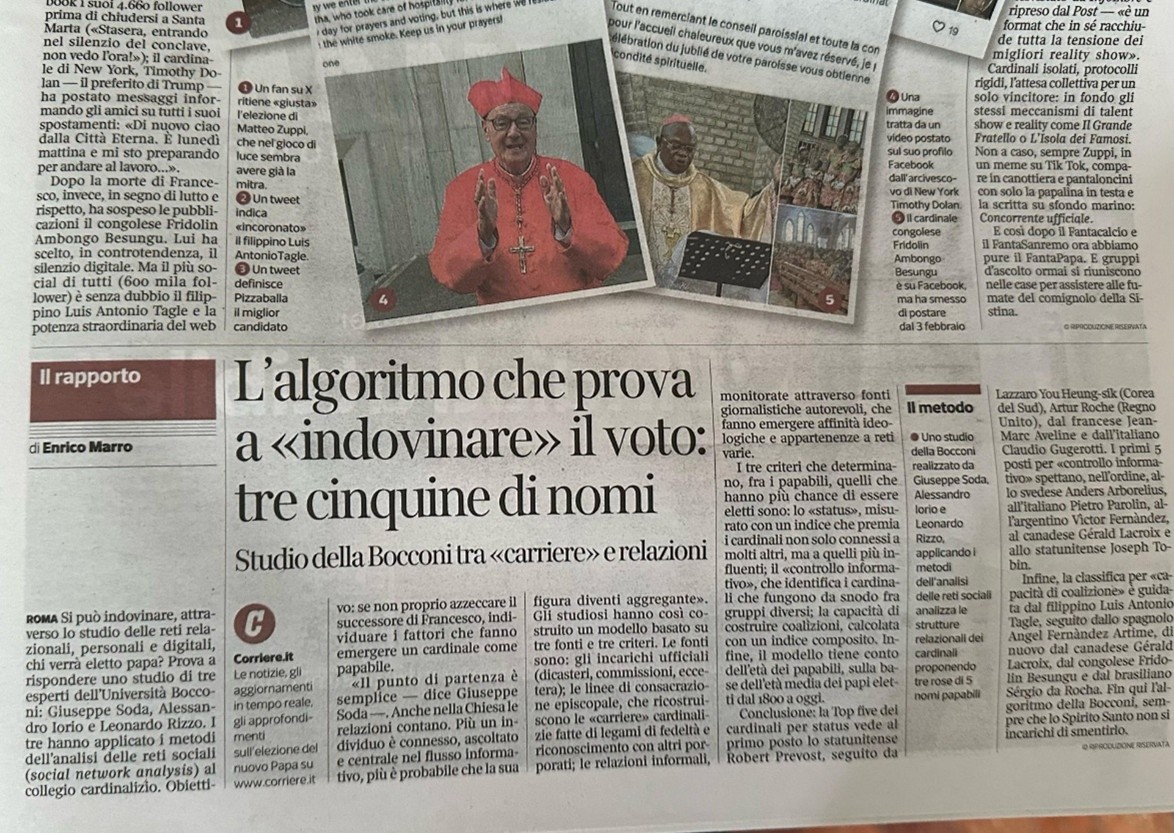}
\par\end{centering}
\centering{}\caption{Printed copy of Corriere della Sera (p. 11: “The algorithm that tries to “guess” the vote: Three sets of five names) available across Italy 19 hours before the Pope election.}\label{fig:corriere}
\end{figure}

As mentioned, this data collection was based only on top figures of the Roman Curia and included a historical view of comembership in collegial bodies, assuming ties persist over time. As we explain, this analysis also incorporats a small, selected number of purely informal ties based on interpersonal bonds and relationships, as identified through journalistic reports that we accessed. These informal ties play a minimal role in the College of Cardinals network given their limited number.

\begin{figure}[h]
\begin{centering}
\includegraphics[width=0.98\textwidth]{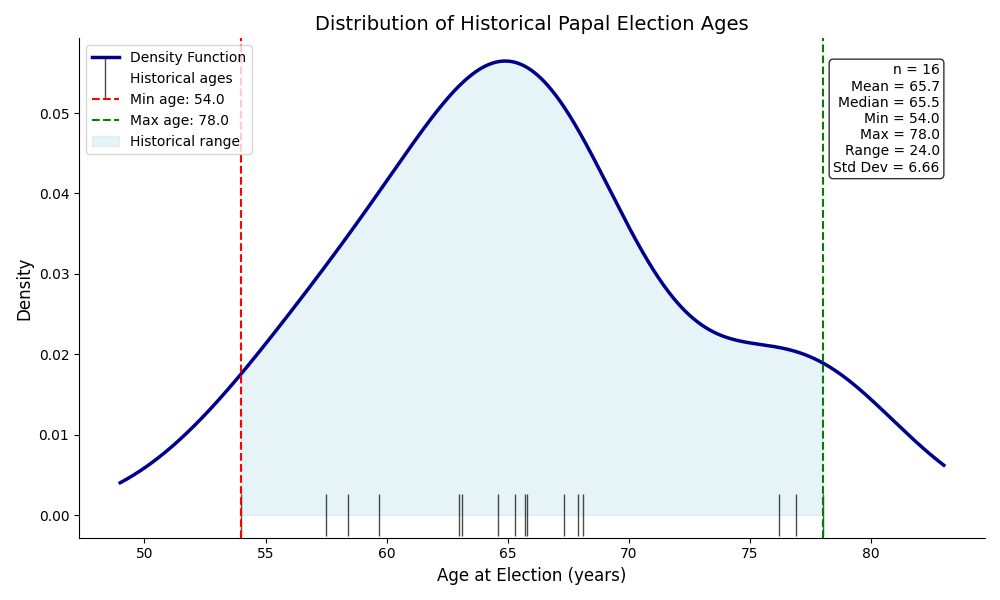}
\par\end{centering}
\centering{}\caption{Probability density function of historical papal election age}\label{fig:popekde}
\end{figure}

\paragraph{Age Adjustment}
\label{sec:age-adj}
We also performed some age adjustment procedures, which provide a statistical framework for assessing papal candidates by integrating network measures with historical age patterns. This methodology employs kernel density estimation to model historical precedent to generate adjusted scores. The framework begins with a dataset of ages from previous papal elections, encompassing 16 historical cases dating back to the 19th century, with recorded ages ranging from 54 to 78 years at the time of election. This historical distribution is analyzed using non-parametric kernel density estimation (KDE), which creates a continuous probability density function representing the likelihood of election based on candidate age (figure \ref{fig:popekde})

The procedure combines two distinct distributions:
\begin{enumerate}
    \item Initial network score: Derived from normalized candidate scores that represent network metrics.
    \item Age-based score: Generated by evaluating each candidate's age against the historical KDE model, then normalizing these density values.
\end{enumerate}

These distributions are integrated using a weighted average approach, controlled by an adjustable parameter that determines the relative influence of age factors versus other qualifications. This weighting mechanism allows for calibration based on the perceived importance of historical age patterns in the selection process.

\paragraph{Rankings}
We report in Table \ref{tab:old} the three rankings for status, information control, and coalition building.

\begin{table}[h]
\centering{}
\caption{Online Ranking}
\begin{tabular}{cccc}
\textbf{\#} & \textbf{Status} & {\bfseries{}%
\begin{tabular}{c}
\textbf{Information}\tabularnewline
\textbf{Control}\tabularnewline
\end{tabular}} & {\bfseries{}%
\begin{tabular}{c}
\textbf{Coalition}\tabularnewline
\textbf{Building}\tabularnewline
\end{tabular}}\tabularnewline
\hline 
 &  &  & \tabularnewline
\textbf{1.} & %
\begin{tabular}{c}
R. Prevost\tabularnewline
Moderate\tabularnewline
US\tabularnewline
\end{tabular} & %
\begin{tabular}{c}
A. Arborelius\tabularnewline
Soft Conservative\tabularnewline
Sweden\tabularnewline
\end{tabular} & %
\begin{tabular}{c}
L. A. Tagle\tabularnewline
Soft Liberal\tabularnewline
Philippines\tabularnewline
\end{tabular}\tabularnewline
 &  &  & \tabularnewline
\textbf{2.} & %
\begin{tabular}{c}
L. You Heung-sik\tabularnewline
Soft Liberal\tabularnewline
South Korea\tabularnewline
\end{tabular} & %
\begin{tabular}{c}
P. Parolin\tabularnewline
Liberal\tabularnewline
Italy\tabularnewline
\end{tabular} & %
\begin{tabular}{c}
A. Fernández Artime\tabularnewline
Soft Liberal\tabularnewline
Spain\tabularnewline
\end{tabular}\tabularnewline
 &  &  & \tabularnewline
\textbf{3.} & %
\begin{tabular}{c}
A. Roche\tabularnewline
Liberal\tabularnewline
UK\tabularnewline
\end{tabular} & %
\begin{tabular}{c}
V. Fernández\tabularnewline
Liberal\tabularnewline
Argentina\tabularnewline
\end{tabular} & %
\begin{tabular}{c}
G. Lacroix\tabularnewline
Moderate\tabularnewline
Canada\tabularnewline
\end{tabular}\tabularnewline
 &  &  & \tabularnewline
\textbf{4.} & %
\begin{tabular}{c}
J.M Aveline\tabularnewline
Soft Liberal\tabularnewline
France\tabularnewline
\end{tabular} & %
\begin{tabular}{c}
G. Lacroix\tabularnewline
Moderate\tabularnewline
Canada\tabularnewline
\end{tabular} & %
\begin{tabular}{c}
F. A. Besungu\tabularnewline
Soft Conservative\tabularnewline
Congo\tabularnewline
\end{tabular}\tabularnewline
 &  &  & \tabularnewline
\textbf{5.} & %
\begin{tabular}{c}
C. Gugerotti\tabularnewline
Soft Liberal\tabularnewline
Italy\tabularnewline
\end{tabular} & %
\begin{tabular}{c}
J. Tobin\tabularnewline
Liberal\tabularnewline
USA\tabularnewline
\end{tabular} & %
\begin{tabular}{c}
S. da Rocha\tabularnewline
Soft Liberal\tabularnewline
Brazil\tabularnewline
\end{tabular}\tabularnewline
\end{tabular}
\label{tab:old}
\end{table}

\begin{figure}[h]
\begin{centering}
\includegraphics[width=0.95\textwidth]{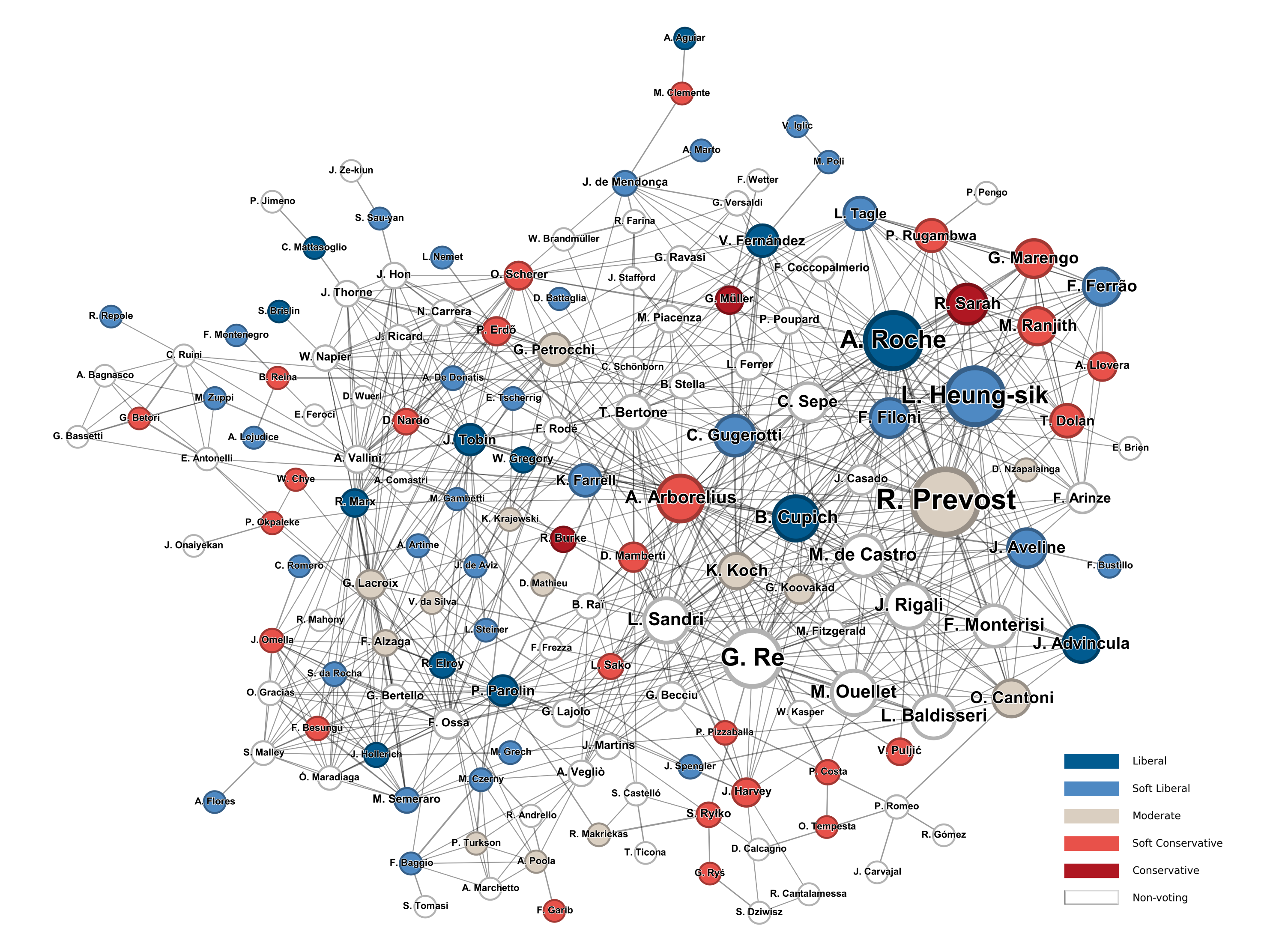}
\par\end{centering}
\centering{}\caption{Multiplex College of Cardinals Network –Doctrinal Orientation}\label{fig:old1}
Nodes are colored by doctrinal orientation (white = non-voting; light blue = soft liberal; dark blue = liberal; grey = moderate; light red = soft conservative; dark red = conservative) and sized by weighted eigenvector centrality.
\end{figure}

\begin{figure}[h]
\begin{centering}
\includegraphics[width=0.95\textwidth]{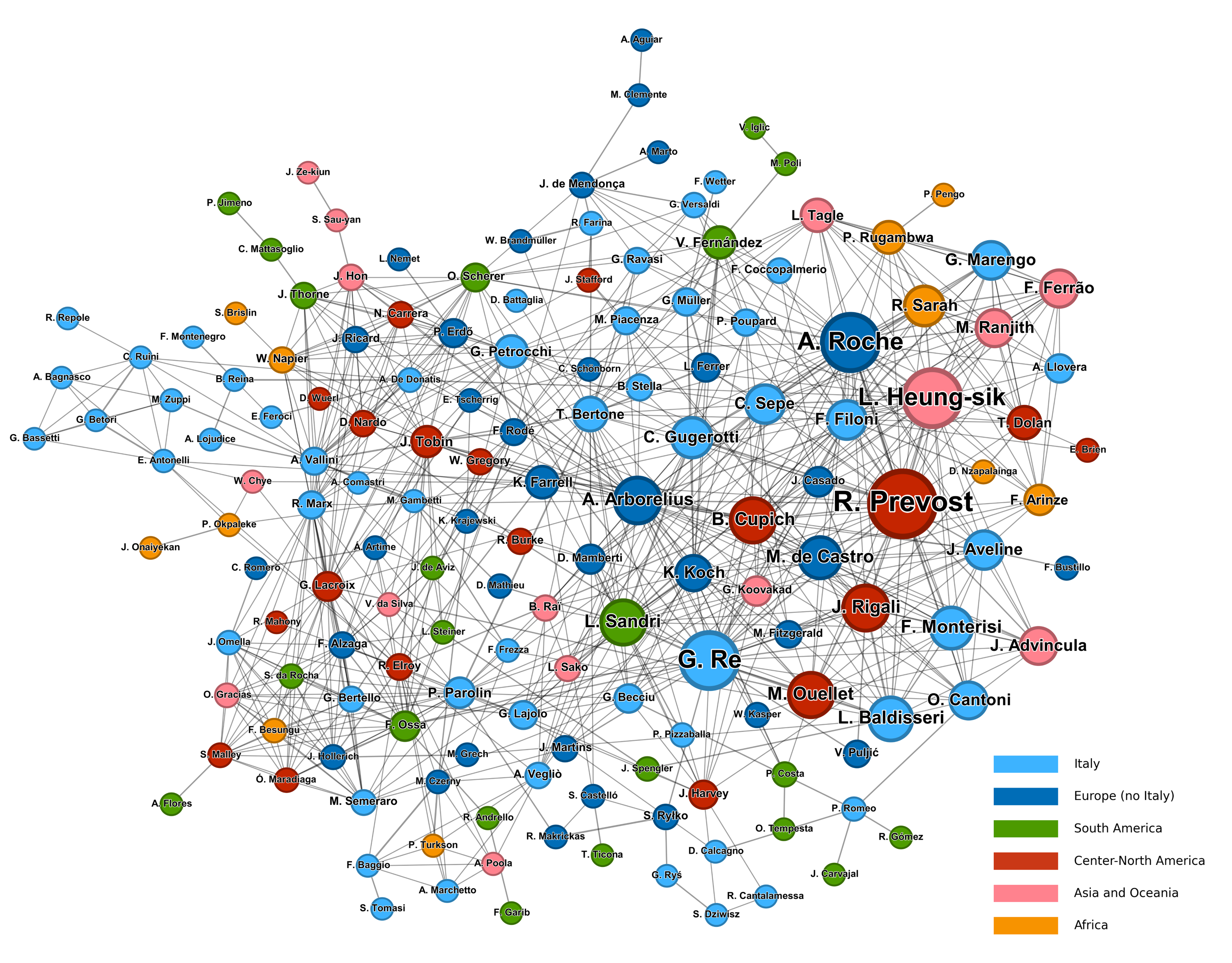}
\par\end{centering}
\centering{}\caption{Multiplex College of Cardinals Network – Geographical Split}
\label{fig:old2}Nodes are colored according to each cardinal’s geographical area of origin and sized by weighted eigenvector centrality.
\end{figure}

\paragraph{Network Visualizations}
The resulting network consisted of 156 nodes and 750 weighted edges, enabling a clear visualization of key patterns of influence and association, while remaining accessible and interpretable for a broader audience. Figures \ref{fig:old1} and \ref{fig:old2} reports these networks.

\end{document}